\documentclass[]{pasj01}

\def\OI{{O}\emissiontype{I }}
\def\OVII{{O}\emissiontype{VII }}
\def\OVIII{{O}\emissiontype{VIII }}

\begin{document} 
\Received{}
\Accepted{}

\title{Search for WHIM around A2744 using Suzaku}

\author{Shiho \textsc{Hattori}\altaffilmark{1}}
\author{Naomi \textsc{Ota}\altaffilmark{1}}
\author{Yu-Ying \textsc{Zhang}\altaffilmark{2}}
\author{Hiroki \textsc{Akamatsu}\altaffilmark{3}}
\author{Alexis \textsc{Finoguenov}\altaffilmark{4,5}}

\altaffiltext{1}{Department of Physics, Nara Women's University, Kitauoyanishi-machi, Nara, Nara 630-8506, Japan}
\altaffiltext{2}{Argelander Institute for Astronomy, Bonn University, Auf dem H\"{u}gel 71, 53121 Bonn, Germany}
\altaffiltext{3}{SRON Netherlands Institute for Space Research, Sorbonnelaan 2, 3584 CA Utrecht, The Netherlands}
\altaffiltext{4}{Max-Planck-Institut f\"{u}r  extraterrestrische Physik, Giessenbachstra\ss e, 85748 Garching, Germany}
\altaffiltext{5}{Department of Physics, University of Helsinki, Gustaf H\"{a}llstr\"{o}min katu 2a, FI-00014 Helsinki, Finland}

\KeyWords{cosmology: observations --- large-scale structure of universe --- intergalactic medium 
--- galaxies: clusters: individual (Abell 2744) --- X-rays: galaxies: clusters} 

\maketitle

\begin{abstract}
  We present the results from the Suzaku satellite of the surrounding
  region of a galaxy cluster, A2744, at $z=0.3$. To search for oxygen
  emission lines from the warm-hot intergalactic medium (WHIM), we
  analyzed X-ray spectra from two northeast regions 2.2--3.3 and
  3.3--4.4~Mpc from the center of the cluster, which offers the first
  test on the presence of a WHIM near the typical accretion shock
  radius ($\sim 2r_{200}$) predicted by hydrodynamical
  simulations. For the 2.2--3.3~Mpc region, the spectral fit
  significantly ($99.2$\% significance) improved when we include
  \OVII and \OVIII lines in the spectral model. 
A comparable WHIM surface brightness was
  obtained in the 3.3--4.4~Mpc region and the redshift of \OVIII line
  is consistent with $z=0.3$ within errors. The present results
  support that the observed soft X-ray emission originated from the
  WHIM.  However, considering both statistical and systematic
  uncertainties, \OVIII detection in the northeast regions was
  marginal.  The surface brightness of the \OVIII line in ${\rm
      10^{-7} photons\,cm^{-2}s^{-1}arcmin^{-2}}$ was measured to be
    $2.7\pm 1.0$, $2.1 \pm 1.2$ for the 2.2--3.3, 3.3--4.4~Mpc
    regions, giving the upper limit on the baryon overdensity of
    $\delta = 319 (< 443) $, $283 (<446)$, respectively.  This is
  comparable with previous observations of cluster outskirts and their
  theoretical predictions. The future prospect for WHIM detection
  using the Athena X-IFU micro-calorimeter is briefly discussed
  here. In addition, we also derived the ICM temperature distribution
  of A2744 to detect a clear discontinuity at the location of the
  radio relic. This suggests that the cluster has undergone strong
  shock heating by mass accretion along the filament.
\end{abstract}

\section{Introduction}
In the present Universe, the baryon mass falls short of the most
likely value predicted by Big Bang nucleosynthesis by a factor of 2;
this is called the ``missing baryon problem''
\citep{1998ApJ...503..518F}. Numerical simulations predict that a
large fraction of baryons should be in the form of warm-hot
intergalactic medium (WHIM) with temperature $10^{5} - 10^{7}$~K and
have a very diffuse distribution \citep{1999ApJ...514....1C}.  The
baryon overdensity, a ratio between the hydrogen density and its mean
value in the Universe, $\delta \equiv n_{\rm H}/\bar{n_{\rm H}}$, is
predicted to be $10 - 500$ for the WHIM \citep{1999ApJ...514....1C,
  2012ApJ...759...23S}. The detection of this missing warm-hot gas is
not only a key to resolving the discrepancy of the total amount of
baryons between the nearby and distant universe but also to understand
their evolution.

Identification of the WHIM is a challenge given its very low surface
brightness and the current sensitivities of the available instruments
(see \cite{2007ARA&A..45..221B} for a review). At $10^{5} - 10^{7}$~K,
emission or absorption lines from ionized atoms may be detectable in
the soft X-ray and ultraviolet regimes. In particular, oxygen is the
most abundant heavy element, X-ray spectroscopy of \OVII and \OVIII
lines can then provide an important clue in the search for the
WHIM. Since the WHIM is expected to be most dense around clusters of
galaxies, the analysis of soft X-ray emission in and near such objects
has been carried out.

Possible detections of redshifted \OVII line emission in 7 clusters of
galaxies have been reported by
\citet{2003A&A...397..445K,2004JKAS...37..375K} based on XMM-Newton
observations. \citet{2007PASJ...59S.339T} observed the surrounding
regions of A2218 at $z=0.18$ using Suzaku and derived an upper limit
of the \OVII and \OVIII fluxes that is significantly lower than
previous measurements, yielding an overdensity of $\delta <
270$. Using XMM-Newton, \citet{2008A&A...482L..29W} detected X-ray
emission from the filament connecting the double cluster A222/A223,
which corresponds to $\delta \sim 150$, while
\citet{2012PASJ...64...18M} found no significant WHIM emission in a
region between A3558 and A3556, giving an upper limit of $\delta <
380$. Recently, based on long XMM-Newton observations,
\citet{2015Natur.528..105E} discovered hot gas that coincides with the
overdensities of galaxies and dark matter in the filaments around
A2744, suggesting that a large fraction of the cosmic baryons are
located in the cosmic web.

A2744 (AC118 or RXC J0014.3--3022; \cite{2004A&A...425..367B}) is an
active merging galaxy cluster at $z=0.308$ with a complex mass
structure and is also known as ``Pandora's cluster'' (e.g.,
\cite{2011ApJ...728...27O}). The precise lensing mass distribution has
been determined by utilizing the deepest HST data from the Frontier
Fields program \citep{2015MNRAS.452.1437J,2015ApJ...811...29W}.
Large-scale galaxy filaments in the south and northwest directions
were first noted by \citet{2007A&A...470..425B}. Spectroscopically
identified 343 member galaxies within 3~Mpc from the cluster center
were catalogued by \citet{2011ApJ...728...27O}, with which a galaxy
concentration in the east-northeast was also recognized. Using Suzaku,
\citet{2014A&A...562A..11I} measured the intracluster medium (ICM)
temperature distribution in the three directions out to a virial
radius of $r_{200}=2$~Mpc and found that there was not a clear
temperature decrease in the cluster's outskirts, unlike many other
clusters \citep{2013SSRv..177..195R}. They also pointed out the
presence of high-temperature gas in the northeast region whose
location coincides with a large radio relic
\citep{2007A&A...467..943O}, suggesting that the cluster has
experienced shock heating due to a merger or mass accretion.
 
In this paper, we will analyze X-ray spectra taken from deep Suzaku
observations of the A2744 northeast region with the aim of studying
the WHIM emission as well as the thermodynamical properties of the
ICM. Since this follow-up observation covered a wider area in
comparison with the previous X-ray observations, we can not only
examine the WHIM emission reported by XMM-Newton, but can also extend
the measurements out to 4.4~Mpc from the cluster's center. This will
provide the first test on the presence of warm-hot gas near $r = 2
r_{200}$, which corresponds to the typical radius where accretion
shock occurs \citep{2015ApJ...806...68L}. The X-ray Imaging
Spectrometer (XIS; \cite{2007PASJ...59S..23K}) onboard Suzaku
\citep{2007PASJ...59S...1M} has the lowest background level
available. In addition to the improved energy resolution in the soft
X-ray band, it is suited for studying faint diffuse emission from
warm/hot gas beyond the cluster virial radius. To complement the
spatial resolution of Suzaku's X-ray telescopes, XMM-Newton data were
utilized to eliminate contamination from point sources in the field.

Throughout this paper, we will adopt the following cosmological
parameters, $H_0 = 70~{\rm km\,s^{-1}\,Mpc^{-1}}$, $\Omega_M = 0.27$,
and $\Omega_{\Lambda} = 0.73$. $1\arcmin = 274~{\rm kpc}$ at a cluster
redshift of $z=0.308$ (The NASA/IPAC Extragalactic Database). The
quoted errors are at the $1\sigma$ confidence level unless otherwise
stated. The solar abundance table by \citet{1989GeCoA..53..197A} was
assumed.

\section{Observation and data reduction}\label{sec:obs}
We conducted a follow-up observation of the A2744 northeast filament
using Suzaku for 83~ksec and performed data analysis in combination
with the previous central and southern pointing
observations. Table~\ref{tab:obslog} provides the observation log. The
data taken using two front-side illuminated CCDs (XIS-0, XIS-3) and
one backside illuminated CCD (XIS-1) were reduced in the standard
manner by using {\tt HEASoft} version 6.17 and a calibration database
released on 2015 October 5. 

Since we are interested in soft X-ray emission from the intergalactic
medium, we have to pay attention to possible contamination from
neutral oxygen lines produced in the Earth's atmosphere
\citep{2014PASJ...66L...3S, 2013PASJ...65...32Y}.  The spectral
analysis was performed with changing the data screening criteria of
the minimum elevation angle from the limb of the day Earth, ${\rm
  DYE\_ELV}$, from $20\degree$ to $50\degree$. The \OI line flux
significantly reduced if the threshold was changed from 20\degree to
30\degree, and statistically consistent fitting results were obtained
if ${\rm DYE\_ELV} > 30\degree$. Thus we show results based on the
data selection with ${\rm DYE\_ELV} > 30\degree$ in the present
paper. Systematic uncertainties in the WHIM measurement are examined
in \S\ref{subsec:whim_sb} in more detail.

The CCD flickering pixels were removed according to a recipe provided
by the XIS instrument
team\footnote{http://www.astro.isas.jaxa.jp/suzaku/analysis/xis/nxb\_new2/}. During
the Suzaku observations of A2744 Northeast, XIS light curves were
stable and the solar-wind proton flux measured by the ACE satellite,
$<2\times10^{8}~{\rm cm^{-2}s^{-1}}$, agree with that of the quiescent
phase \citep{2007PASJ...59S.133F}. Thus the present data is unlikely
to be affected by the solar activity.

\begin{table*}
\tbl{Log of Suzaku observations}{%
\begin{tabular}{lllll}  \hline\hline
Object & OBSID & Date & (RA, Dec) & Exposure (s)$^{\mathrm{a}}$  \\ \hline
A2744 Center & 802033010 & 2007 May 19--23 & (00:14:9.5, -30:20:40.6) & 153 241 \\
A2744 South  & 805015010 & 2010 Dec. 10--12 & (00:14:3.2, -30:33:2.9) & 69 776 \\
A2744 Northeast     & 808008010 & 2013 Nov. 20--22 & (00:14:52.5, -30:18:28.1) & 82 781 \\ \hline
\end{tabular}}\label{tab:obslog}
\begin{tabnote}
$^{\mathrm{a}}$ Exposure time after data filtering.
\end{tabnote}
\end{table*}

\begin{figure}
\begin{center}
 \includegraphics[width=8cm]{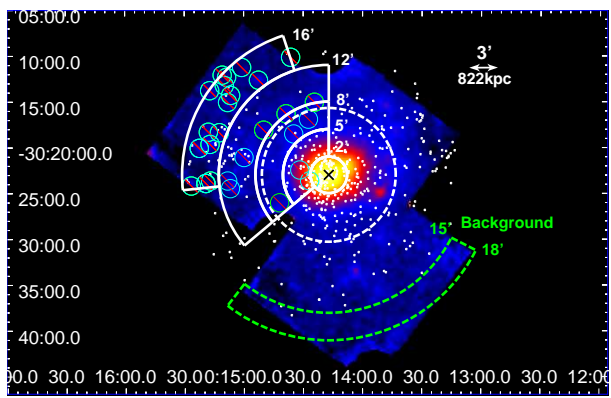}
 \end{center}
 \caption{Suzaku XIS-0 + XIS-1 + XIS-3 mosaic image of A2744 in the
   0.5--10~keV band.  The background is not subtracted but the image
   is corrected for the exposure map and vignetting effects. The X-ray
   centroid of the cluster emission is marked by ``$\times$''. The
   white sectors are the spectral integration regions for ICM and WHIM
   analyses, while the green sector was used to model background
   spectra. The cyan circles denote point sources detected by
   XMM-Newton, which were excluded with an $r=1'$ radius from the
   spectral integration regions. The virial radius is indicated by the
   dashed white circle. The positions of member galaxies with
   redshifts within $cz \pm 5000~{\rm km\,s^{-1}}$ ($z=0.308$) and the
   $r_{\rm F}$-band magnitude $r_{\rm F}<23$
   \citep{2011ApJ...728...27O} are marked by the white
   dots.}\label{fig:Simage}
\end{figure}

The XIS mosaic image in the 0.5--10~keV band is shown in
Figure~\ref{fig:Simage}. As the virial radius is estimated as $r_{200}
= 2.0$~Mpc or $7\arcmin.3$ \citep{2014A&A...562A..11I}, the northeast
pointing covers a large radial range that is $0-2.2r_{200}$ from the
cluster's center. To search for diffuse WHIM emission beyond the
cluster's virial radius, two sectors at $8\arcmin-12\arcmin$ and
$12\arcmin-16\arcmin$ were defined as filament regions because the
northeastern galaxy concentration at around $r \sim 10\arcmin$ was
noted by the galaxy density map
\citep{2011ApJ...728...27O,2014A&A...562A..11I}.
To measure the ICM temperature profile across the radio relic
at $r=7'$, additional three integration regions, $0\arcmin-2\arcmin$,
$2\arcmin-5\arcmin$, $5\arcmin-8\arcmin$ were defined as indicated in
the figure. 

The XIS spectra in each region were extracted from the cleaned XIS
event files.  Since the three regions at $0\arcmin-2\arcmin$,
$2\arcmin-5\arcmin$, $5\arcmin-8\arcmin$ were covered by both Center
and Northeast observations, the spectra extracted from these two data
sets were co-added. The spectra were rebinned so that each spectral
bin contained more than 20 source counts. The X-ray telescope and CCD
responses were calculated using {\tt xissimarfgen} and {\tt
  xisrmfgen}, respectively.  Non-X-ray background produced by {\tt
  xisnxbgen} was subtracted from the spectra. \ \ Point sources in the
XIS fields were searched by utilizing the XMM-Newton PN and MOS data
(OBSID 0743850101), and detected sources whose fluxes are larger than
$2\times10^{-14}~{\rm erg\,s^{-1}cm^{-2}}$ were removed from the XIS
data with an $r=1\arcmin$ circle. In the case that the extraction
radius was changed from $1\arcmin$ to $1\arcmin.5$, we confirmed that
the present results were not significantly affected.

\begin{figure}[hbt]
\begin{center}
 \includegraphics[width=8cm]{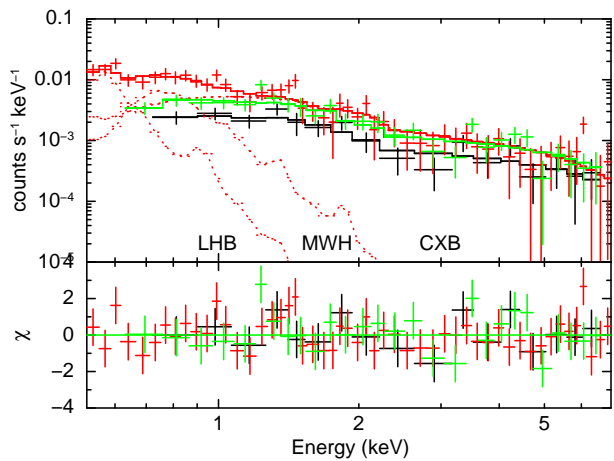}
 \end{center}
 \caption{XIS background spectra with subtracted non-X-ray background
   component. XIS-0, XIS-1 and XIS-3 are shown using black, red, and
   green crosses, respectively. The solid and dashed lines represent
   the total model and the background components for XIS-1,
   respectively.}\label{fig:bgdS}
\end{figure}

We derived the background model by fitting the spectra in the southern
region ($15\arcmin < r < 18\arcmin$) to the same model as used by
\citet{2014A&A...562A..11I}. The model function is composed of the
cosmic X-ray background (CXB) and the galactic X-ray background (GXB)
from the Milky Way halo (MWH) and the Local Hot Bubble (LHB) and is
represented by ``apec$_{\rm LHB}$+phabs(apec$_{\rm
  MWH}$+power-law$_{\rm CXB}$)''. Here, the MWH and LHB temperatures
were fixed at 0.34 and 0.13~keV, respectively, and the power-law index
of the CXB component was fixed at 1.4. The hydrogen column density of
the galactic absorption was fixed at $N_{\rm H} = 1.39\times
10^{20}~{\rm cm^{-2}}$ \citep{2005A&A...440..775K}. The fitting result
is shown in Figure~\ref{fig:bgdS}. The model parameters are listed in
Table~\ref{tb:bgds}, which agree with those obtained from previous
Suzaku analysis \citep{2014A&A...562A..11I} and XMM-Newton
observations within their errors \citep{2015Natur.528..105E}.

We also analyzed two offset XIS fields located at about $3\degree$
away from the A2744 center (OBSID: 506028010, 506029010) to find their
CXB and GXB models are consistent with those presented in
Table~\ref{tb:bgds} and their best-fit parameters agree within
10\%. Therefore we decided to take $\pm10$\% as the systematic error
on the background model.

\begin{table*}
\tbl{Background model parameters}{
\begin{tabular}{llllllll} \hline\hline
 $\Gamma$ & $Norm$$^{\mathrm{a}}$ & $kT_{\rm MWH}$ & $Norm$$^{\mathrm{b}}$ & $kT_{\rm LHB}$ & $Norm$$^{\mathrm{b}}$ & $\chi^2/{\rm d.o.f.}$ \\ 
                & $(\times10^{-4})$ & $[{\rm keV}]$ & $(\times10^{-4})$ & [{\rm keV}] & $(\times10^{-3})$ & & \\ \hline
1.4(fixed) & 6.58$^{+0.54}_{-0.54}$ & 0.34(fixed) & 3.35$^{+0.60}_{-0.56}$ & 0.13(fixed) & 1.71$^{+0.35}_{-0.32}$ & 77.5/92 \\  \hline
\end{tabular}} \label{tb:bgds}
\begin{tabnote}
$^{\mathrm{a}}$Normalization of power-law model [photons keV$^{-1}$~cm $^{-2}$ s$^{-1}$].
$^{\mathrm{b}}$Normalization of APEC model, $Norm = \int n_{\rm e}n_{\rm H}dV/(4\pi(1+z)^2D^2_A)$~[10$^{-14}$~cm$^{-5}]$. $D_A$ is the angular diameter distance to the source. An $r=20'$ uniform sky is assumed.
 \end{tabnote}
\end{table*}

\section{Search for WHIM emission}\label{sec:analysis_whim}
\subsection{Analysis of WHIM emission spectra}\label{subsec:analysis_whim}
To search for WHIM emission in the northeast of A2744, we analyzed the
XIS spectra from the $r = 8\arcmin -12\arcmin (=1.1r_{200} - 1.6
r_{200})$ and $r=12\arcmin -16\arcmin (=1.6r_{200}-2.2r_{200})$
regions (Figure~\ref{fig:Simage}). This enabled us to examine the
presence of WHIM near $r = 2 r_{200}$, corresponding to the typical
location of the accretion shock as suggested by hydrodynamical
simulations \citep{2015ApJ...806...68L}. In this analysis, only XIS-1
data were used because it has a higher sensitivity in the soft X-ray
band than other XIS sensors.

\begin{figure*}
\begin{center}
\rotatebox{0}{\scalebox{0.25}{\includegraphics{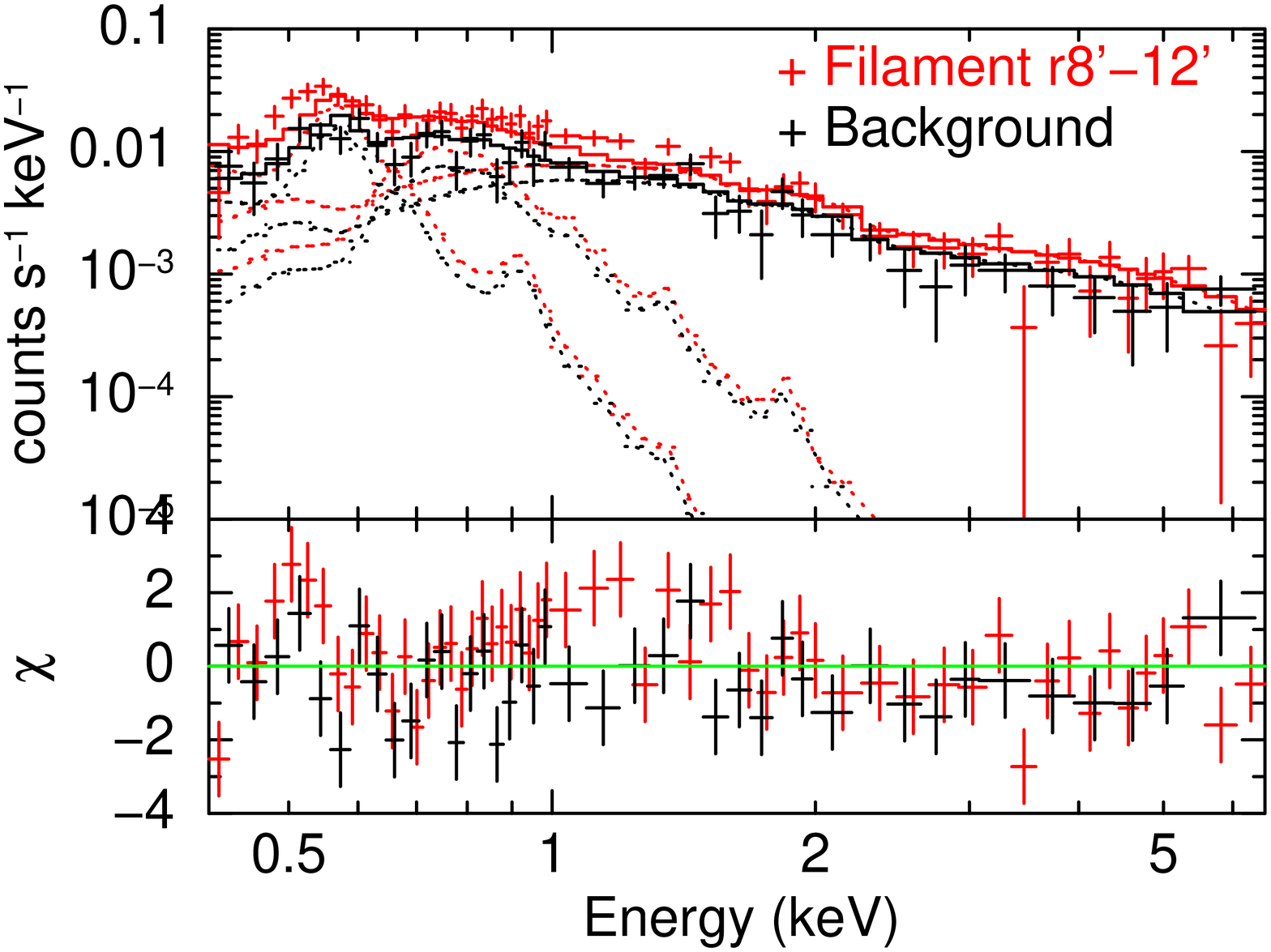}}}
\rotatebox{0}{\scalebox{0.25}{\includegraphics{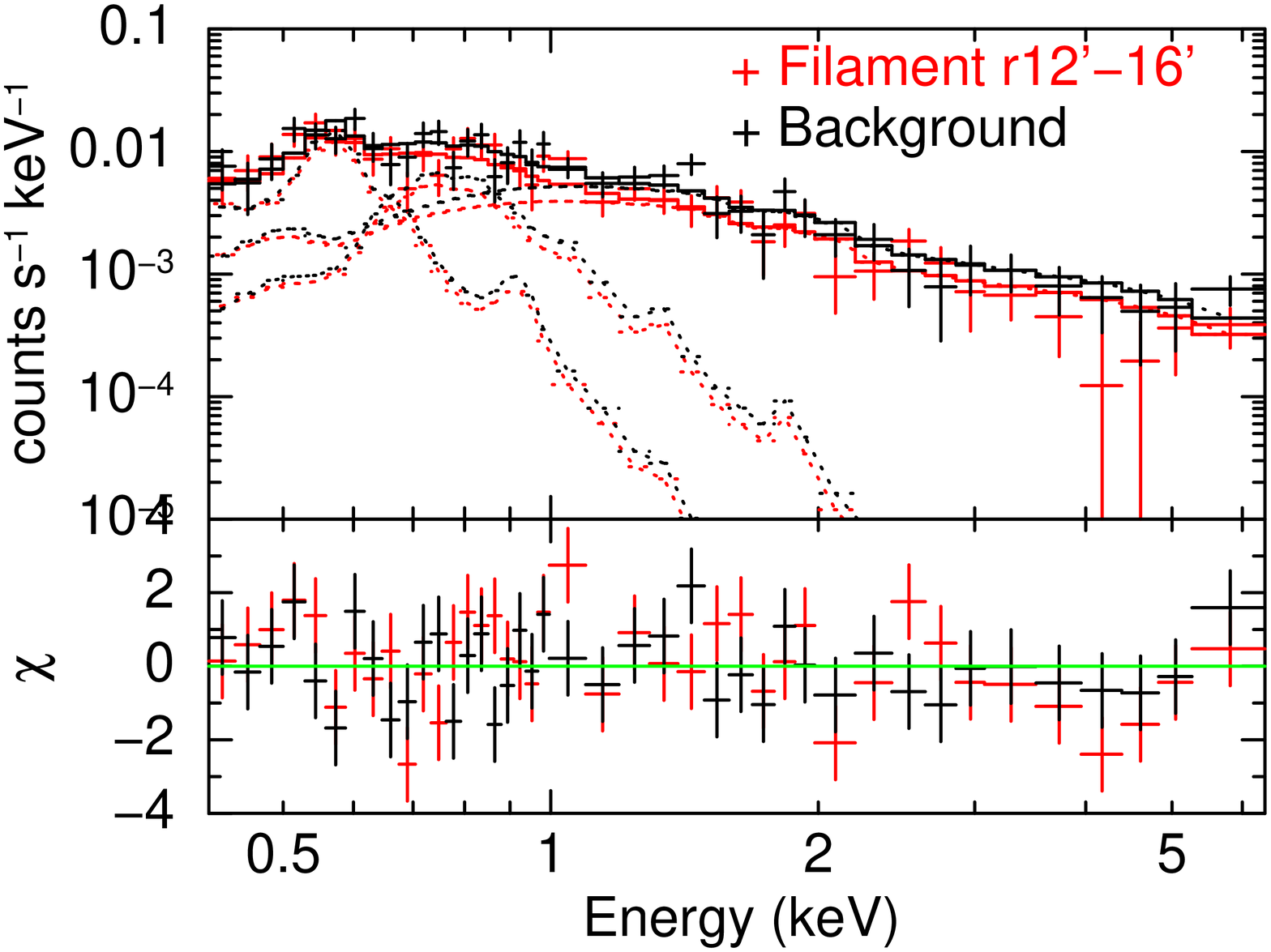}}}

\rotatebox{0}{\scalebox{0.25}{\includegraphics{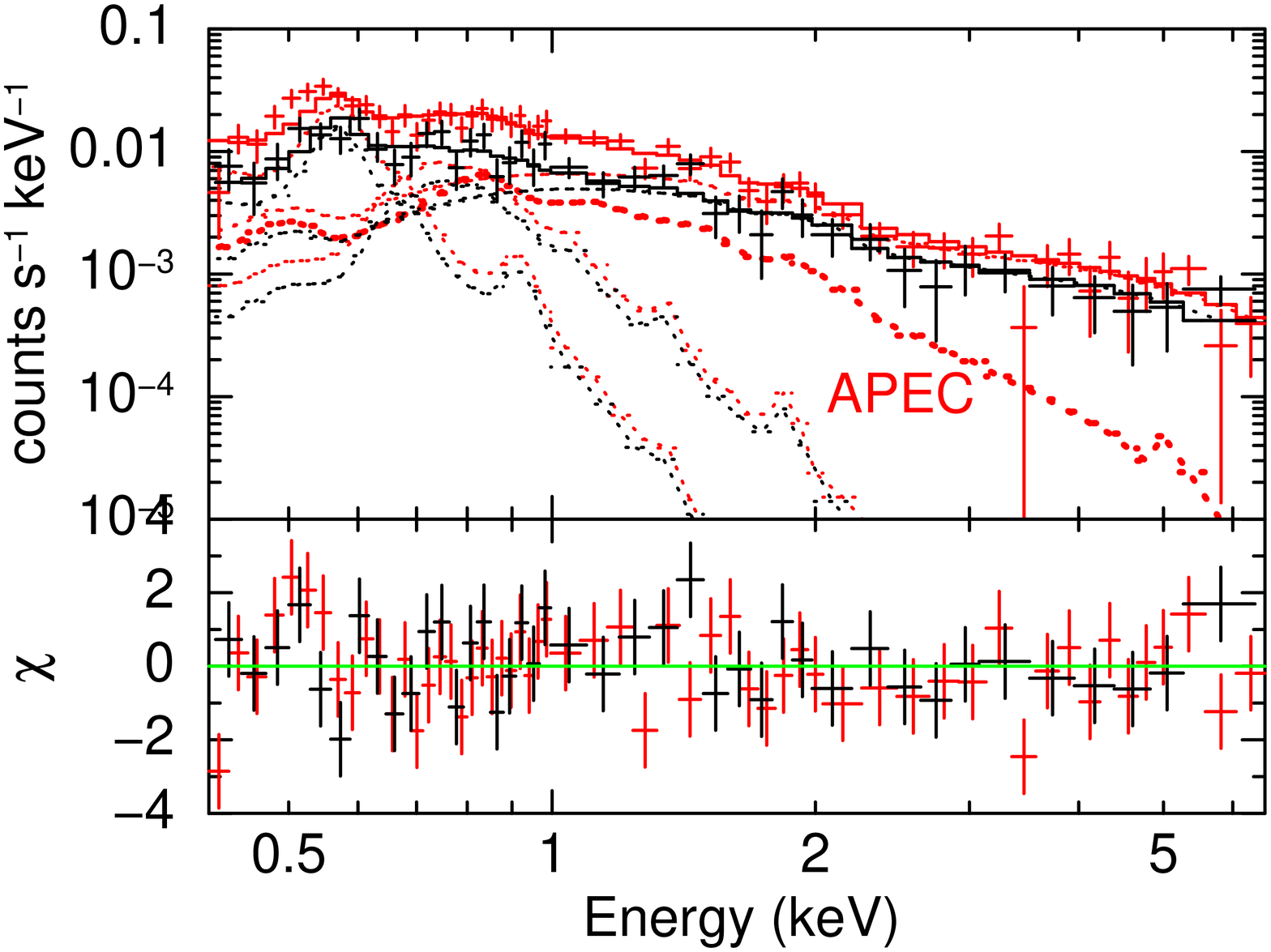}}}
\rotatebox{0}{\scalebox{0.25}{\includegraphics{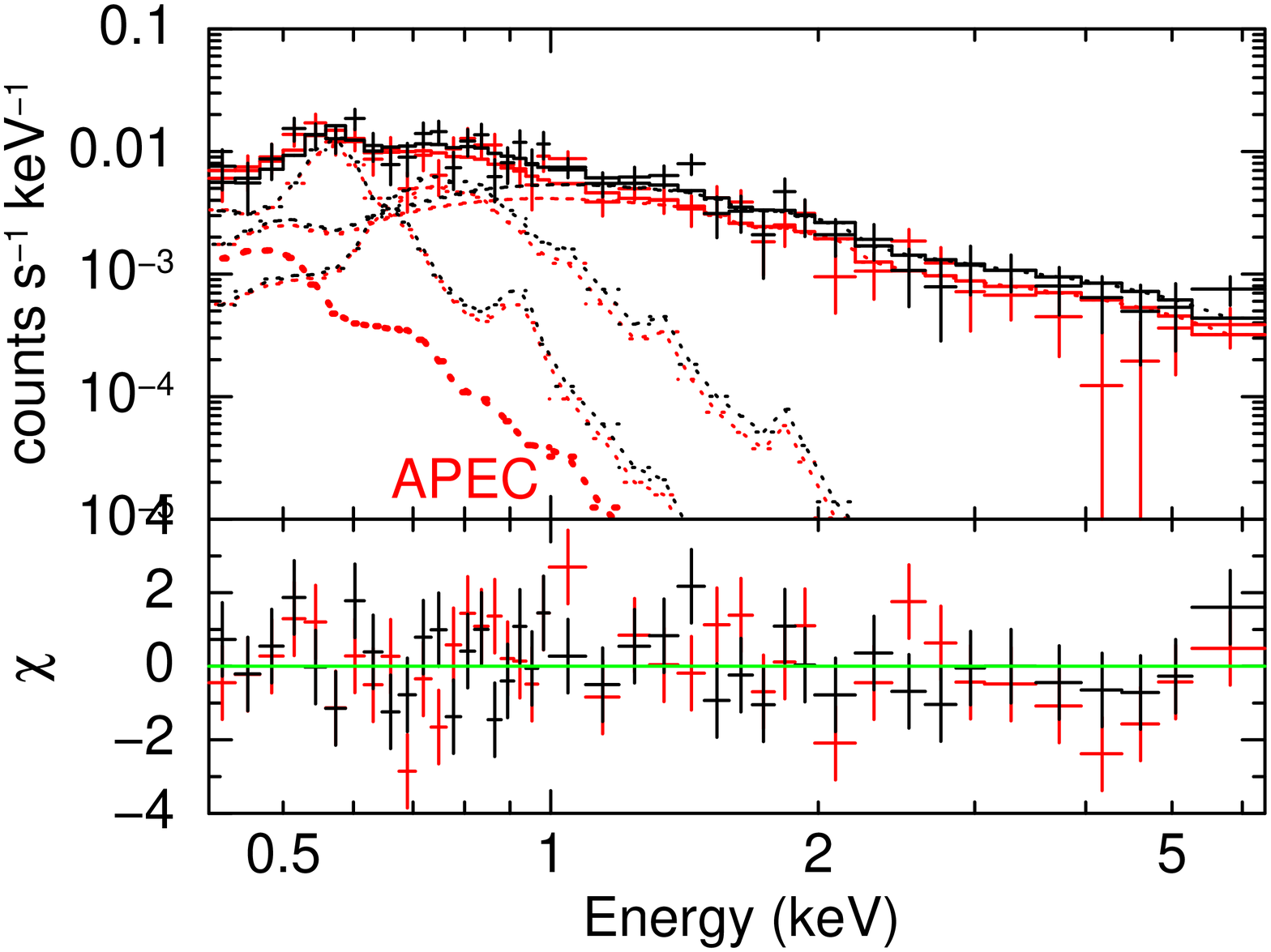}}}

\rotatebox{0}{\scalebox{0.25}{\includegraphics{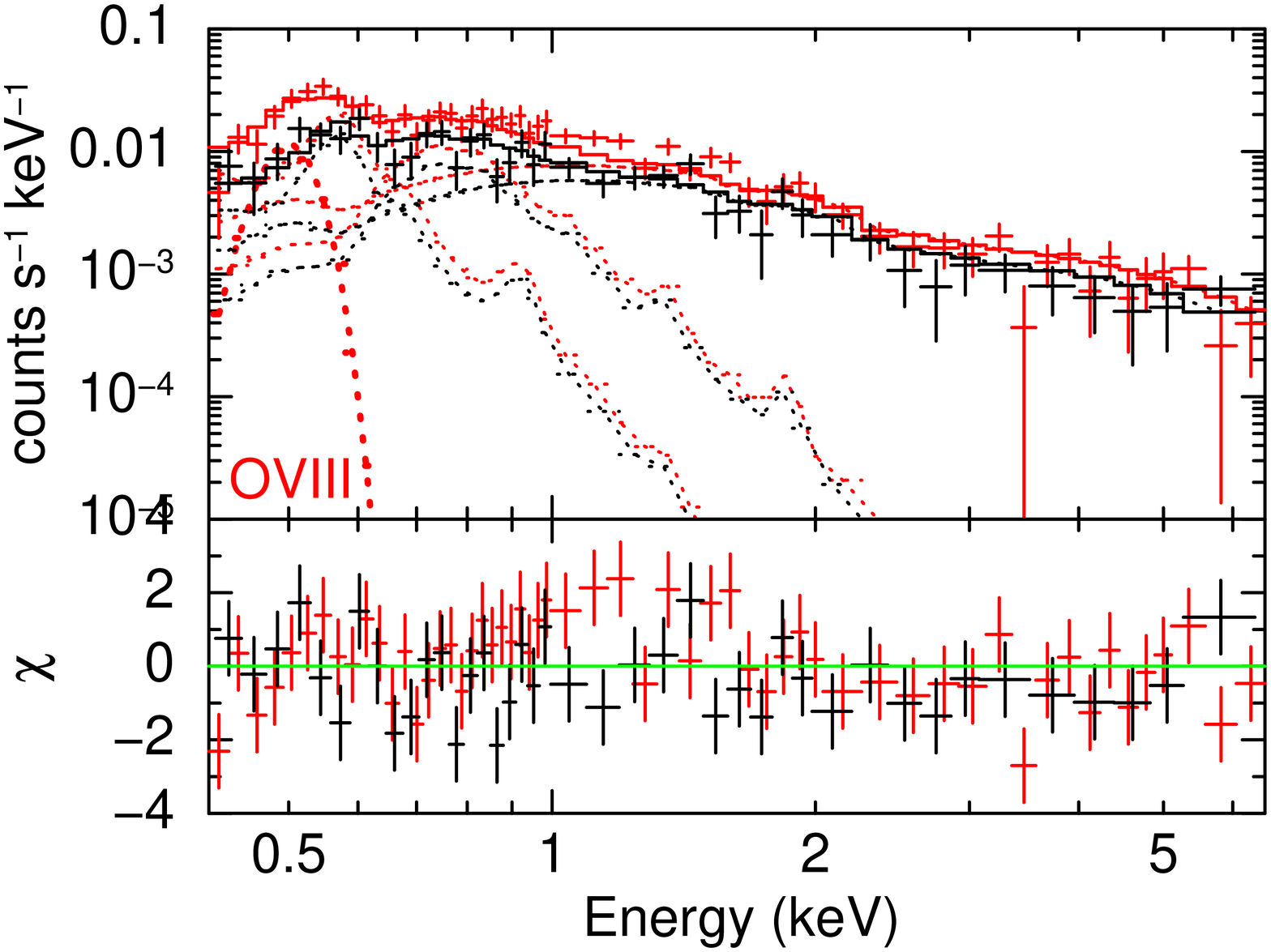}}}
\rotatebox{0}{\scalebox{0.25}{\includegraphics{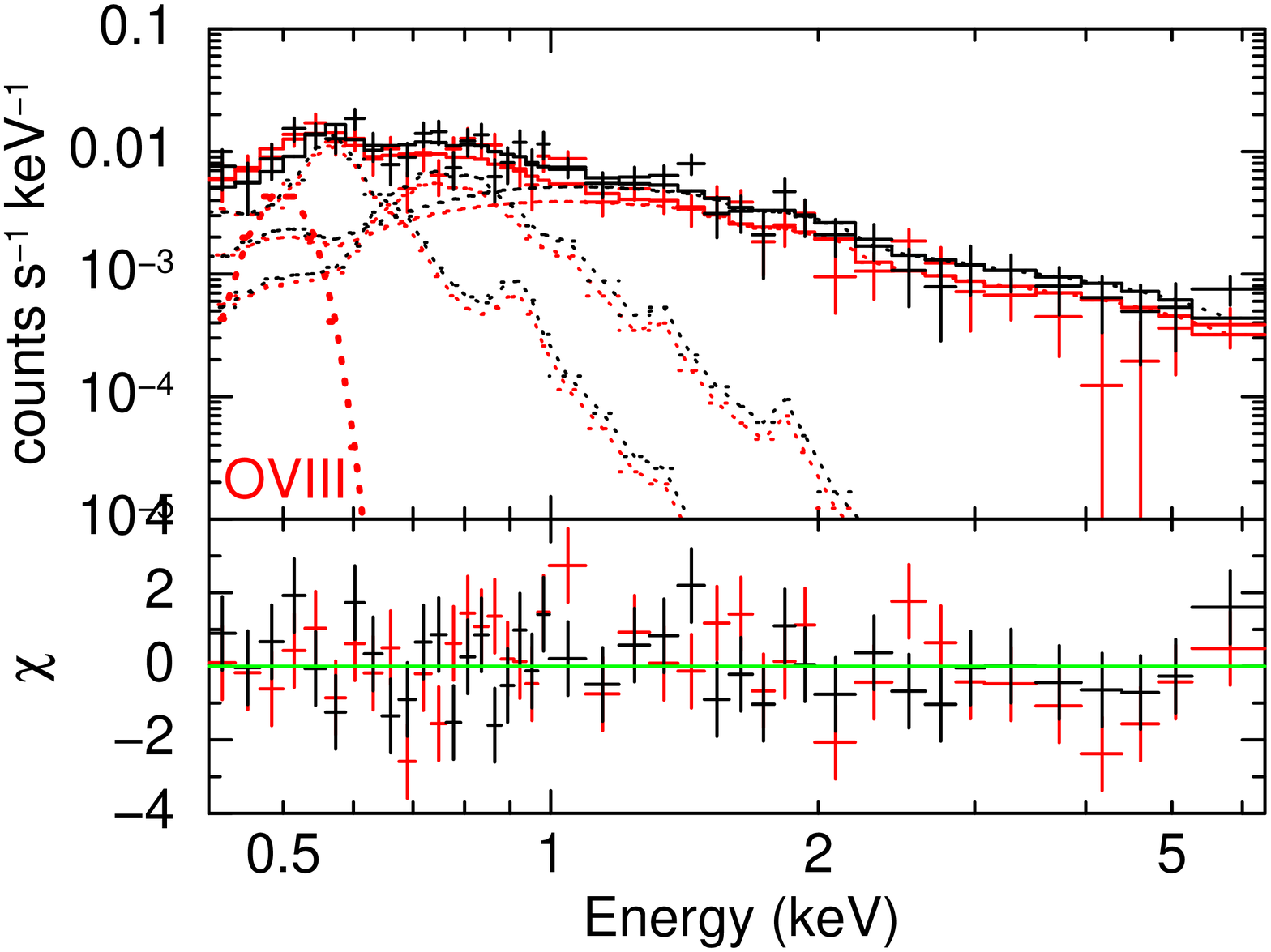}}}
 \end{center}
 \caption{Left: XIS-1 spectra of the A2744 north-east filament
   $8\arcmin<r<12\arcmin$ (red) and the background region
   (black). Right: XIS-1 spectra of the A2744 north-east filament
   $12\arcmin<r<16\arcmin$ (red) and the background region (black). In
   the upper panels, the total model (the solid line) is composed of
   GXB and CXB (thin dotted lines). In the middle panels, the total
   model incorporates i) the additional APEC model 
   for the WHIM (thick dotted line). In the bottom panels, the total
   model incorporates ii) the additional two Gaussian lines for the
   WHIM (thick dotted line).  }\label{fig:apecS}
\end{figure*}

First, the WHIM emission in the filament spectrum was modeled in two
ways: i) the APEC thin-thermal plasma model
\citep{2001ApJ...556L..91S} and ii) two Gaussian functions for the
redshifted \OVII and \OVIII lines.  In case i), the metal abundance
was fixed at $Z=0.2$~solar because the median oxygen metallicity is
0.18~solar \citep{2006ApJ...650..560C}. In the outermost region
($12\arcmin < r < 16\arcmin$), the temperature was not well
constrained and fixed at the typical value of $kT_1=0.2$~keV since the
\OVIII line is the strongest at this temperature. In case ii), the
line centroid energies were fixed at 0.439~keV and 0.500~keV and the
line width was fixed at $\sigma = 0$. Since the WHIM emission is
faint, we fitted the filament and background spectra in the 0.4--7~keV
band simultaneously assuming that the background model is common
between the two regions instead of subtracting it directly from the
filament data. Except for the three normalization factors, the
background model parameters were fixed at the values shown in
Table~\ref{tb:bgds}.  The $\chi^2$ fitting was performed using {\tt
  XSPEC} version 12.9.0\footnote{XSPEC version 12.9.0o was used when
  we analyzed the WHIM component under the APEC model with $z\neq0$.}.
The XIS spectra are shown in Figure~\ref{fig:apecS}, and model
parameters for i) the APEC model and ii) the Gaussian model are shown
in Table~\ref{tb:specfitS}.

\begin{table*}
\begin{center}
\tbl{Model parameters for the A2744 northeast filament}{
\begin{tabular}{lllllllll} \hline\hline
Model &      & \multicolumn{2}{c}{APEC$^{\mathrm{a}}$} & \multicolumn{2}{c}{APEC$^{\mathrm{a}}$}  & Gaussian, \OVIII$^{\mathrm{b}}$ &Gaussian, \OVII$^{\mathrm{b}}$  &  \\ \cline{3-4}  \cline{5-6} \cline{7-8}
& $r$ & $kT_1$ &  $Norm_1$ & $kT_2$ &  $Norm_2$ & $Norm^{\mathrm{c}}$ & $Norm^{\mathrm{c}}$ & $\chi^2$/d.o.f. \\  
& $[\rm arcmin]$ & [keV] & ($\times10^{-3}$) & [keV] & ($\times10^{-3}$)  &  ($\times10^{-4}$)  & ($\times10^{-4}$)  & \\ \hline 
i)   & 8--12 & $1.73^{+0.35}_{-0.15}$ &$2.85^{+0.49}_{-0.50}$ & -- & -- & -- & -- & 97.3/93 \\
i)   &12--16 & 0.2 (fixed) & $1.51^{+1.32}_{-1.32}$ & -- & -- & -- & -- & 92.5/78 \\ \hline
ii)  & 8--12 & -- & -- & -- & -- &  $4.22^{+1.05}_{-1.05}$ & $<0.39$ & 113.9/93 \\
ii)  & 12--16 & -- &  -- & -- & -- & $2.65^{+1.28}_{-1.28}$ & $<1.65$ & 89.5/77 \\ \hline
iii) & 8--12 & $1.73^{+0.50}_{-0.16}$ & $2.60^{+0.48}_{-0.51}$ & -- &  --  & $3.35^{+1.04}_{-1.10}$  &  $<0.32$ & 87.5/91 \\ \hline
iv) & 8--12 & $1.74^{+0.47}_{-0.15}$ & $2.55^{+0.55}_{-0.51}$ & $0.27^{+0.09}_{-0.05}$ & $1.45^{+1.01}_{-1.01}$ & --  &  -- & 95.6/91 \\ \hline
\end{tabular}}\label{tb:specfitS}
\end{center}
\begin{tabnote}
$^{\mathrm{a}}${The metal abundance was fixed at 0.2~solar.}
$^{\mathrm{b}}${Centroid energies of the redshifted emission lines were assumed to be 0.500~keV and 0.439~keV for the \OVIII and \OVII lines, respectively. Their rest-frame energies are 0.654 and 0.574 ~keV.} 
$^{\mathrm{c}}$Gaussian line flux~${\rm [photons\,cm^{-2}s^{-1}]}$.
\end{tabnote}
\end{table*}

In comparison with the fit using the background-only model
($\chi^2$/d.o.f. = 130.0/95 and 93.8/79 for $r=8\arcmin - 12\arcmin$
and $12\arcmin - 16\arcmin$, respectively), we examined the
improvement of the spectral fit by adding the WHIM model using the
F-test (Table~\ref{tb:ftest}). The F-statistic values were i) 15.7 and
ii) 6.6 for 2 additional model parameters. Thus the WHIM component is
statistically significant at the $>99.99$\% and 99.8\% levels. 

\begin{table*}
\begin{center}
\tbl{Surface brightness of the WHIM and the statistical significance by F-test}{
\begin{tabular}{llllllll} \hline \hline
$r$[arcmin] & Model component& $I^{\mathrm{a}}$ & $I$($2\sigma$ limit)$^{\mathrm{b}}$& $F$$^{\mathrm{c}}$ & Probability$^{\mathrm{c}}$ &$\delta$$^{\mathrm{d}}$ &$\delta$ ($2\sigma$ limit)$^{\mathrm{e}}$\\ \hline
8--12 & i) APEC & $6.67\pm1.14\pm1.94$  & $<11.2$ & 15.7 & $1.4\times10^{-6}$ & $199\pm17\pm26$ & $<261$ \\  \hline   
8--12 & ii) Gaussian, \OVIII  & $3.36\pm0.84\pm0.70$ & $<5.5$ & 6.60 & 0.002 & $358 \pm 44 \pm 38$ & $<474$ \\ 
         & Gaussian, \OVII  & $0.00\pm0.31\pm0.00$ & $<0.6$ & & & $0 \pm 93 \pm 0$ & $<132$ \\ \hline  
8--12 & iii) Gaussian, \OVIII  & $2.67\pm0.85\pm0.60$ & $<4.8$ & 5.08 & 0.008 & $319 \pm 50 \pm 36$ & $<443$ \\ 
         & Gaussian, \OVII  & $0.00\pm0.25\pm0.00$ & $<0.5$ & & & $0 \pm 85 \pm 0 $ & $<120$ \\ \hline  
8--12 & iv) APEC & $2.57\pm1.79\pm2.15$  & $<8.2 $ & 0.81& 0.45 & $166\pm58\pm70$ & $<348$ \\  \hline   
12--16 & i) APEC ($kT_1$ fix) & $2.08\pm1.82\pm1.67$  & $<7.0$ & 1.10 & 0.30 & $145\pm63\pm60$ & $<319$ \\ \hline 
12--16 & ii) Gaussian, \OVIII  & $2.11\pm1.02\pm0.65$ & $<4.5$ & 1.86 & 0.16 & $283 \pm 68 \pm 44$ & $<446$ \\  
          & Gaussian, \OVII  & $0.00\pm1.31\pm0.00$ & $<2.6$ & & & $0 \pm 193 \pm 0$& $<272$ \\ \hline   
\end{tabular}}\label{tb:ftest}
\end{center}
\begin{tabnote}
$^{\mathrm{a}}$The surface brightness of the WHIM component in units of $10^{-7}{\rm photons~cm^{-2}~s^{-1}arcmin^{-2}}$. The first and second errors are the $1\sigma$ statistical and systematic uncertainties. The brightness was calculated in the 0.4--7~keV band for the APEC model.
$^{\mathrm{b}}$The $2\sigma$ upper limit on the surface brightness of the WHIM component. Both statistical and systematic errors were taken into account. 
$^{\mathrm{c}}$The F-statistic value and the probability of finding a higher F-value than observed (see text). 
$^{\mathrm{d}}$ The baryon overdensity and the the $1\sigma$ statistical and systematic uncertainties. 
$^{\mathrm{e}}$ The $2\sigma$ upper limit on the baryon overdensity.
\end{tabnote}
\end{table*}

\begin{figure}[hbt]
\begin{center}
\rotatebox{0}{\scalebox{0.25}{\includegraphics{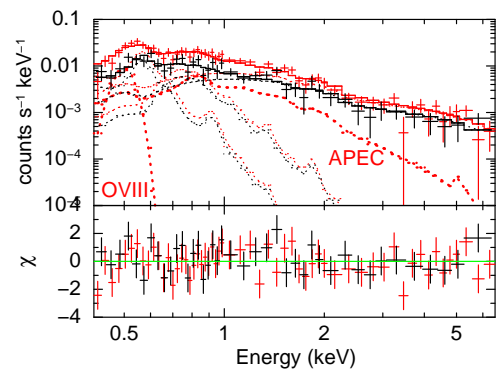}}}
 \end{center}
 \caption{XIS-1 spectrum of the A2744 north-east filament
   $8\arcmin<r<12\arcmin$ (red) and the background region (black). The
   total model (the solid line) is composed of GXB+CXB (thin dotted
   lines) and iii) the APEC model and Gaussian lines (thick dotted
   lines).}\label{fig:specfitS}
\end{figure}

 We should note that the gas temperature at $r=8\arcmin-12\arcmin$
  derived under i) the APEC model, $kT_1=1.7^{+0.35}_{-0.15}$~keV, is
  higher than that predicted for the WHIM \citep{1999ApJ...514....1C}
  and positive residuals are seen at around 0.5~keV, while ii) the
  Gaussian model left positive residuals in consecutive data points at
  1--2 keV. Thus, in order to test a possibility that the observed spectrum
  is represented by a superposition of ICM and WHIM emissions, we fit
  the XIS-1 spectra by assuming iii) a two-component model consisting
  of an APEC model for the ICM and two Gaussian lines for the
  WHIM. The result is shown in Table~\ref{tb:specfitS} and
  Figure~\ref{fig:specfitS}. In comparison with the case i), the
  improvement of the spectral fit by adding the WHIM component is
  significant at 99.2\% by F-test. We also confirmed that a reasonable
  fit is obtained if the Gaussian lines were replaced by a
  $Z=0.2$~solar APEC model, resulting the best-fit WHIM temperature of
  $kT_2=0.27^{+0.09}_{-0.05}$~keV and $\chi^2/{\rm d.o.f.}=95.6/91$
  (model iv in Table~\ref{tb:specfitS}). Therefore we suggest that the
  observed soft X-ray spectra can be well reproduced by a sum of
  $kT_1\sim1.7$~keV ICM and $kT_2\sim0.2-0.4$~keV WHIM.

For $r=12\arcmin - 16\arcmin$, i) and ii) provided a reasonably good
fit to the data. In comparison with the fit using the background-only
model, the F-values were i) 1.1 for one additional parameter and ii)
1.9 for two additional parameters and the improvement was not
significant at the 95\% level.
 
If the redshift of oxygen lines was allowed to vary, the spectral fit
gave the centroid energy of the redshifted \OVIII line $E_0 =
0.522\pm0.009\pm0.005$, $0.514\pm0.017\pm0.005$~keV or $z =
0.253\pm0.023\pm0.012$, $0.272\pm0.044\pm 0.012$ for
$r=8\arcmin-12\arcmin$, $12\arcmin-16\arcmin$.  Here the first and
second errors represent the $1\sigma$ statistical and systematic
uncertainties. The latter was estimated by fitting the \OI line
spectra taken at $10\degree<{\rm DYE\_ELV} < 20\degree$ to the
Gaussian line. Since the observed \OI line energy is found to be
systematically higher by 5~eV, we assign the $1\sigma$ systematic
error of 5~eV to the line energy \footnote{The typical calibration
  error of the XIS energy scale in the soft X-ray band is 5~eV
  (http://www.astro.isas.jaxa.jp/suzaku/process/caveats/)}.  Therefore
the fitted redshifts are marginally lower than the cluster redshift
$z=0.308$ but agree with the mean optical redshift of the northeast
filament region, $z=0.3$, \citep{2011ApJ...728...27O} within
$1.8\sigma$, $0.6\sigma$ for the two regions, respectively. This
further supports that the soft X-ray emission in the filament region
originates from the WHIM at $z=0.3$.

Note that we repeated the analysis by fitting unbinned XIS spectra
with the C-statistics and confirmed that the above results on the
surface brightness and the centroid energy of the \OVIII line were not
affected by the choice of fitting methods.

\subsection{WHIM surface brightness and systematic uncertainties}\label{subsec:whim_sb}
The surface brightness of the WHIM emission, $I$, derived from the
spectral analysis is listed in Table~\ref{tb:ftest}, where the first
and second errors refer to the $1\sigma$ statistical and systematic
uncertainties. The latter was estimated in the following manner.

We first consider two possible sources of systematic errors: a)
uncertainty in the background modeling and b) degradation of XIS's
low-energy efficiency due to contaminating material on the optical
blocking filter.  We then examine c) possible contamination from the
neutral oxygen lines.

For a), since the GXB also has strong oxygen lines, we intentionally
changed the GXB intensity by $\pm 10$\% (see \S\ref{sec:obs}). For b),
by comparing a measured thickness of the contaminant with a model used
in the response generation (The Suzaku Technical Description Sec 7),
the uncertainties in absorption column densities of carbon, nitrogen,
and oxygen are estimated to be $\pm 10$\%. To take this into account,
the absorption column densities were included in the spectral
model. The total systematic error in the surface brightness was then
calculated by adding the above two errors in quadrature.

As a result, for $r=8\arcmin-12\arcmin$, the WHIM brightness in
  ${\rm photons\,cm^{-2}s^{-1}arcmin^{-2}}$ was obtained as ii) $I=(3.36\pm0.84\pm0.70)\times 10^{-7}$ and iii)
  $I=(2.67\pm0.85\pm0.60)\times 10^{-7}$ for the \OVIII line model. 
  Thus, the \OVIII emission is statistically significant at the
  $3.1-4.0\sigma$ level.  For $r=12\arcmin-16\arcmin$, though the WHIM
brightness derived under the two models, i)
$I=(2.08\pm1.82\pm1.67)\times10^{-7}$ and ii)
$I=(2.11\pm1.02\pm0.65)\times10^{-7}$, is comparable to that in the
$r=8\arcmin-12\arcmin$ region, the statistical significance of the
\OVIII line was lower ($2.1\sigma$). Hence the detection of the WHIM
component is marginal if systematic uncertainty is considered.

For c), as mentioned in \S\ref{sec:obs}, the fitting results for the
WHIM component were not affected by the data screening criteria if
${\rm DYE\_ELV}>30\degree$ was applied. This suggests that the present
data is less contaminated by the \OI line emission. Given the current
instrumental resolution and photon statistics, however, the line
centroid energy derived from the Gaussian fitting indicates that the
\OI line at 0.525~keV cannot be completely distinguished from the
redshifted \OVIII line at 0.500~keV. For example, if the \OVIII line
was replaced with the \OI line at 0.525~keV, we obtained an acceptable
fit to the $r=12\arcmin-16\arcmin$ spectra ($\chi^2/{\rm d.o.f.} =
89/78$).

To check how much the WHIM surface brightness is affected by the
possible \OI contamination, we added another Gaussian line at
0.525~keV to  iii) the APEC and Gaussian models for
  $r=8\arcmin-12\arcmin$ and ii) the Gaussian model for
$12\arcmin-16\arcmin$(\S\ref{subsec:analysis_whim}) and fitted the XIS
spectra again. This yielded a lower \OVIII brightness with a large
statistical error: $I =(0.12\pm0.76\pm0.07)\times 10^{-7}$,
$(0.93\pm1.37\pm0.77)\times 10^{-7}{\rm
  photons~cm^{-2}~s^{-1}arcmin^{-2}}$ for the above regions,
respectively.

\section{Analysis of ICM temperature profile}\label{sec:analysis_icm}
The previous Suzaku analysis suggested the existence of very hot,
shock-heated gas with $kT=13.6^{+9.6}_{-6.7}$~keV (90\% error
including the systematic uncertainty due to point source subtraction)
that coincided with the location of the large radio relic. To improve
the measurement accuracy and examine the shock structure, we derived a
radial profile of the ICM temperature out to $r=12\arcmin$ in the
northeast direction based on a follow-up observation. Note that
  the outermost sector at $r=8\arcmin-12\arcmin$ is common to one of
  the filament regions defined in \S\ref{sec:obs} and analyzed in
  \S\ref{sec:analysis_whim}.

\begin{figure}[hbt]
\begin{center}
 \rotatebox{0}{\scalebox{0.25}{\includegraphics{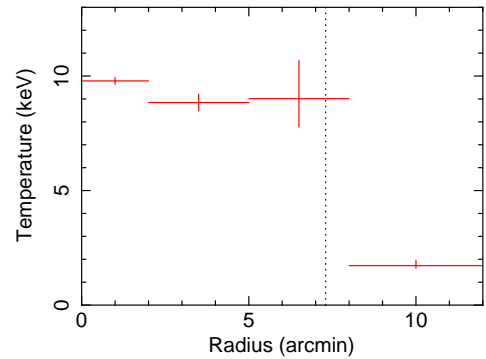}}}
 \end{center}
 \caption{ICM temperature profile in the A2744 northeast region as a
   function of distance from the cluster center (arcmin).  The dashed
   line indicates the virial radius.}\label{fig:temp}
\end{figure}

  The observed 0.7--7~keV spectra of three sensors (XIS-0, XIS-1, and
  XIS-3) were fitted using the APEC model corrected for the Galactic
  absorption. The background in each annular region was simulated by
  assuming the best-fit CXB+GXB parameters (Table~\ref{tb:bgds}) and
  were subtracted from the XIS spectra (\S\ref{sec:analysis_icm}). The
  redshift and Galactic hydrogen column density were fixed at
  $z=0.308$ and $N_{\rm H} = 1.39\times 10^{20}~{\rm cm^{-2}}$
  \citep{2005A&A...440..775K}. The ICM temperature, metal abundance,
  and the normalization factor were treated as free parameters. The
  best-fit model parameters are given in Table~\ref{tb:temp} and the
  temperature profile is plotted in Figure~\ref{fig:temp}. The
  temperature ranged between 8--10~keV inside the virial radius.  The
  temperature of $9.0^{+1.7}_{-1.2}$~keV at $5\arcmin<r<8\arcmin$ is
  statistically consistent with previous results
  \citep{2014A&A...562A..11I} and that of the relic region,
  $12.3^{+4.5}_{-3.5}$~keV, reported based on the joint Suzaku and
  XMM-Newton analysis \citep{2016MNRAS.461.1302E}. On the other hand,
  the temperature is a factor of $5.2\pm1.0$ lower outside that
  radius, i.e., $1.72^{+0.24}_{-0.11}$~keV. This ICM temperature
    agrees well with that obtained from the XIS spectral analysis
    under the model i) or iii) in \S\ref{subsec:analysis_whim} and is comparable to 
   that reported from the XMM-Newton measurement of hot gas at the eastern galaxy
    concentration, $T=(15\pm 2)\times10^{6}$~K \citep{2015Natur.528..105E}.  Our Suzaku
  observation confirms the presence of a clear temperature drop across
  the relic. The nature of this structure is discussed in
  \S\ref{subsec:shock}.

\begin{table}
\begin{center}
\tbl{APEC model parameters for the ICM in the northeast.}{ 
\begin{tabular}{lllll} \hline\hline
 $r$ &$kT$ & Abundance & Norm  & $\chi^2$/d.o.f. \\
 $[{\rm arcmin}]$ & $[{\rm keV}]$ & [solar] & $(\times10^{-3})$ &  \\ \hline
0--2 & $9.79^{+0.15}_{-0.15}$ & $0.23^{+0.02}_{-0.02}$ & $6.63^{+0.04}_{-0.04}$  & 2036.7/2141 \\ 
2--5 & $8.84^{+0.37}_{-0.36}$ & $0.21^{+0.04}_{-0.04}$ & $5.71^{+0.08}_{-0.08}$  & 683.3/681 \\
5--8 & $9.01^{+1.66}_{-1.24}$ & $0.04^{+0.17}_{-0.04}$ & $1.12^{+0.06}_{-0.07}$ & 229.6/241 \\
8--12 & $1.72^{+0.24}_{-0.11}$ & $0.2{\rm(fixed)}$ & $0.49^{+0.09}_{-0.09}$ & 161.2/167 \\ \hline
\end{tabular}} \label{tb:temp}
\end{center}
\end{table}

\section{Discussion}\label{sec:discussion}
We analyzed the X-ray spectra obtained from Suzaku's follow-up
observation of the A2744 northeast filament regions and found that the
soft X-ray spectrum at $r= (1.1-1.6)r_{200}$ is significantly better
fitted by adding the WHIM emission model, although the background
uncertainty is not negligible. We also measured the WHIM emission near
the typical accretion radius ($2r_{200}$) for the first time.  The
redshift of \OVIII line is consistent with $z=0.3$ within errors,
suggesting that the observed soft X-ray emission originated from the
WHIM. The detection was, however, not significant under the present
measurement errors. Thus, we will derive the upper limit of the baryon
overdensity in \S\ref{subsec:overdensity} and will discuss the
prospect of the search for WHIM using the future X-ray mission in
\S\ref{subsec:prospect}.  The ICM temperature profile exhibits a clear
drop at the location of radio relic. We will estimate the Mach number
of the shock and compare it with radio observations in
\S\ref{subsec:shock}.

\subsection{Estimation of the WHIM density}\label{subsec:overdensity}
The baryon overdensity $\delta = n_{\rm H}/ \bar{n_{\rm H}}$ at a
large distance from the A2744 center can be calculated by utilizing
the same equations as used in the study of A2218
\citep{2007PASJ...59S.339T}:
\begin{eqnarray}
I &=& C(T)(1+z)^{-3}n_{\rm e}n_{\rm H}ZL, \\
\bar{n}_{\rm H} &=& 1.77\times10^{-7}(1+z)^3~{\rm cm}^{-3},
\end{eqnarray}
where $I$ is the surface brightness of the warm-hot gas, $C(T)$ is the
line emissivity which depends on the warm-hot gas temperature $T$ and
calculated by the AtomDB database
3.0.2\footnote{http:\/\/www.atomdb.org\/}, $n_e$ is the electron
density, $n_{\rm H}$ is the hydrogen density ($n_{\rm e}/n_{\rm
  H}=1.2$), $Z$ is the metal abundance assuming the solar abundance
table by \citet{1989GeCoA..53..197A}\footnote{The oxygen abundance is assumed to be ${\rm O/H} = 8.51\times 10^{-4}$.}, and $L$ is the path length.
  We adopt $kT=0.2$~keV or $T=2.3\times10^6$~K as assumed in the
  analysis of WHIM spectra, which was also suggested to be reasonable
  based on the two component fit to the $r=8\arcmin-12\arcmin$ spectra
  (\S\ref{subsec:analysis_whim}).  Then substituting
$C(2.3\times10^6$~K), $Z=0.2$~solar, our measurement of the \OVIII
surface brightness under the Gaussian model yields
\begin{eqnarray}
  \delta &=& 319\pm62\,(< 443) \left(\frac{Z}{0.2~{\rm Z_{\odot}}}\right)^{-1/2}\left(\frac{L}{2.5~{\rm Mpc}}\right)^{-1/2}, \label{eq:delta_r8-12}\\ 
  \delta &=& 283\pm81\,(< 446) \left(\frac{Z}{0.2~{\rm Z_{\odot}}}\right)^{-1/2}\left(\frac{L}{2.5~{\rm Mpc}}\right)^{-1/2}, \label{eq:delta_r12-16}\end{eqnarray}
for $r=8\arcmin-12\arcmin$ and $12\arcmin-16\arcmin$, respectively.
In the above calculation, the $1\sigma$ error including both
statistical and systematic uncertainties is quoted and the value in
the parenthesis indicates the $2\sigma$ upper limit.  Note that
$L=2.5$~Mpc was assumed because the spectral integration area of
each filament region is approximated by a square, 2.5~Mpc on a
side.

Similar values can be obtained if the limits on the \OVII line and the
APEC normalization are substituted, suggesting that the measurement is
less affected by the choice of the spectral model.  The results are
summarized in Table~\ref{tb:ftest}.  In \S\ref{subsec:overdensity}, we
considered the impact of possible contamination from the \OI line on
the WHIM measurement. The spectral fitting by ii) the Gaussian model
plus the OI line gave a limit comparable to
Equations~\ref{eq:delta_r8-12}--\ref{eq:delta_r12-16}, i.e., $\delta =
150 (<479)$ and $189(<513)$ respectively for $r=8\arcmin-12\arcmin$
and $12\arcmin-16\arcmin$.

The present result is comparable to that estimated near the virial
radius of A2744 \citep{2015Natur.528..105E} and those in other
clusters \citep{2007PASJ...59S.339T, 2003A&A...397..445K}; and also
agrees with the theoretical prediction by \citet{2006ApJ...650..560C},
which suggested that the baryon overdensity ranges from 1 to 300,
within the measurement errors.

\subsection{Future prospects of the WHIM search}\label{subsec:prospect}
Our Suzaku observations suggest that X-ray spectroscopy of filaments
around galaxy clusters will provide an important clue in the search
for the densest part of the WHIM. By utilizing future non-dispersive,
high-resolution X-ray spectrometer onboard DIOS
\citep{2014SPIE.9144E..2QO} and Athena \citep{2015JPhCS.610a2008B},
emission lines from ionized ions of intergalactic warm-hot gas are
expected to be more clearly detected. The X-IFU micro-calorimeter
onboard Athena will achieve a superior energy resolution of
approximately 2.5~eV \citep{2014SPIE.9144E..2LR, 2016arXiv160400670G}.

\begin{figure}[hbt]
\begin{center}
\includegraphics[width=8cm]{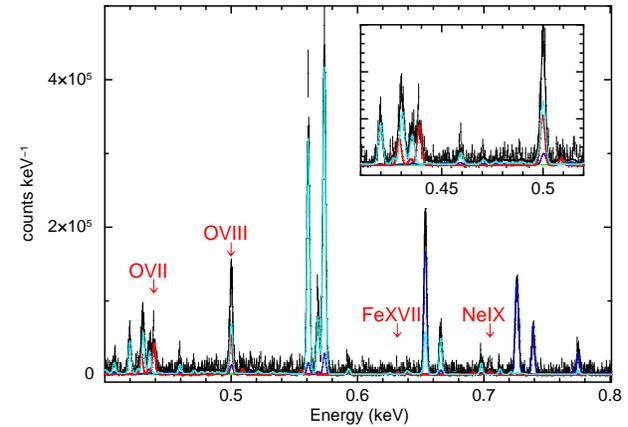}
\end{center}
\caption{Simulated Athena X-IFU spectrum of the WHIM in the case of
  $\delta = 200$ and an exposure time of 20~ks (black cross). The
  spectral components are indicated by different colors: WHIM (red),
  CXB (green), LHB (cyan), and MWH (blue).  Blow-up of the X-IFU
  spectrum in the 0.41--0.52~keV band is also shown.  }\label{fig:sxs}
\end{figure}

To examine the detectability of the WHIM, we simulate the X-IFU
spectra of the A2744 filament using {\tt simx} version 2.4.1
\footnote{RMF (athena\_xifu\_rmf\_v20150327.rmf), ARF
  (athena\_xifu\_1190\_onaxis\_pitch265um\_v20150327.arf) response
  files and the instrumental background file
  (int1arcmin2\_athena\_xifu\_1190\_onaxis\_pitch265um\_v20150327.pha)
  were used.}. The APEC model with $kT=0.2$~keV, $Z=0.2$~solar,
$z=0.3$, and $\delta = 100$, 200, and 400 were assumed. An example of
the X-IFU simulation is shown in Figure~\ref{fig:sxs}. In the case of
$\delta=100$, the count rates of \OVII and \OVIII lines are 175 and
210 (counts/100ks), respectively. Thus if $\delta > 100$, we expect to
detect them with an exposure time of $<10$~ks. Since the centroid
energy of the redshifted \OVIII line overlaps with that of the
Galactic \OVII line, observing \OVII and other lines to identify the
WHIM is important. If the overdensity is high, we can detect not only
the oxygen lines but also Fe and Ne lines. It should be possible to
determine the redshift of the WHIM to an accuracy of about 0.1\%.

\subsection{ICM shock structure at the radio relic}\label{subsec:shock}
In \S\ref{sec:analysis_icm}, we confirmed the presence of a
temperature drop across the radio relic, which suggests that the
cluster has experienced shock heating due to a merger. From the
Ranking-Hugoniot shock condition with the ratio of specific heats
$\gamma=5/3$, the ratio of pre-shock ($T_1$) and post-shock ($T_2$)
temperatures is related to the Mach number $M_{\rm X}$ through the
following relationship \citep{1959flme.book.....L}:
\begin{equation}
\frac{T_2}{T_1} = \frac{5M_{\rm X}^4 + 14 M_{\rm X}^2 -3}{16M_{\rm X}^2}.
\end{equation}
From the observed ICM temperatures, $T_1 = 1.7\pm 0.2$~keV ($8\arcmin
< r < 12\arcmin$) and $T_2 = 9.0\pm 1.5$~keV ($5\arcmin < r <
8\arcmin$), we obtain $M_X = 3.7\pm0.4$. 

From the XMM-Newton observation, an X-ray surface brightness jump was
found at the location of the radio relic \citep{2016MNRAS.461.1302E},
indicating a weak shock with the Mach number of
$1.7^{+0.5}_{-0.3}$. In addition, the spectral index of synchrotron
emission from the radio relic \citep{2007A&A...467..943O} yields a
Mach number of $M_{\rm radio} = 1.7 - 2.5$. The Mach number evaluated
from our X-ray analysis is marginally higher than the above two
measurements, however, we consider that it provides another piece of
evidence that the gas was shock heated via mass accretion along the
filament.

\begin{ack}
  We are grateful to the Suzaku team members for satellite operation
  and instrumental calibration. We also thank the anonymous referee
  and Koji Mori for useful comments.  This work was supported in part
  by JSPS KAKENHI grant 16K05295, 25247028 (NO).  Y.Y.Z. acknowledges
  support by the German BMWi through the Verbundforschung under grant
  50~OR~1506.  H.A. is supported by a Grant-in-Aid for JSPS Fellows
  (26-606).  H.A. acknowledges the support of NWO via a Veni
  grant. This research made use of data obtained from Data ARchives
  and Transmission System (DARTS), provided by Center for
  Science-satellite Operation and Data Archive (C-SODA) at ISAS/JAXA,
  and the NASA/IPAC Extragalactic Database (NED) which is operated by
  the Jet Propulsion Laboratory, California Institute of Technology,
  under contract with the National Aeronautics and Space
  Administration.
  
Dedicated to the memory of Dr. Yu-Ying Zhang, who made a great contribution to this field but sadly passed away at a young age. 
\end{ack}


\end{document}